\documentclass[prx,amsmath,amssymb,nofootinbib,showpacs,twocolumn]{revtex4-2}
\usepackage{textcomp}
\usepackage{overpic}

\usepackage{graphicx}
\usepackage{amsmath,amssymb,epsfig}
    \DeclareMathOperator{\sech}{sech}
    \DeclareMathOperator{\sign}{sign}

\newcommand{\cc}{\mbox{c.c.}}

\newcommand{\dd}{\partial}
\newcommand{\rd}{\mathrm{d}}
\newcommand{\pd}[2]{\frac{\partial #1}{\partial #2}}
\newcommand{\td}[2]{\frac{\rd #1}{\rd #2}}

\newcommand{\curl}{\boldsymbol{\nabla} \times}

\newcommand{\eps}{\epsilon}
\newcommand{\veps}{\varepsilon}

\newcommand{\EE}{\mathcal{E}}
\newcommand{\DD}{\mathcal{D}}
\newcommand{\HH}{\mathcal{H}}
\newcommand{\PP}{\mathcal{P}}

\newcommand{\beq}{\begin{equation}}
\newcommand{\eeq}{\end{equation}}
\newcommand{\non}{\nonumber}

\newcommand{\ii}{i}
\newcommand{\ve}[1]{\mathbf{#1}}

\begin{document}

\title{On the speed of wave packets and the  nonlinear Schr\"odinger equation
}
\author{Gregory Kozyreff}
\affiliation{Optique Nonlin\'eaire Th\'eorique, Universit\'e libre de Bruxelles (U.L.B.), CP 231, Belgium}
\begin{abstract}
The universal theory of weakly nonlinear wave packets given by the nonlinear Schr\"odinger equation is revisited. In the limit where the group and phase velocities are very close together, a multiple scale analysis carried out beyond all orders reveals that a single soliton, bright or dark, can travel at a different speed than the group velocity. In an exponentially small but finite range of parameters, the envelope of the soliton is locked to the rapid oscillations of the carrier wave. Eventually, the dynamics is governed by an equation anologous to that of a pendulum, in which the center of mass of the soliton is subjected to a periodic potential. Consequently, the soliton speed is not constant and generally contains a periodic component. Furthermore, the interaction between two distant solitons can in principle be profoundly altered by the aforementioned effective periodic potential and we conjecture the existence of new bound states. These results are derived on a wide class of wave models and in such a general way that they are believed to be of universal validity.
\end{abstract}
\date{\today}
\maketitle

\section{Introduction}
The nonlinear Schr\"odinger equation (NLSE) has been with us since the end of the 1960's, when it emerged almost simulatenously in optics, hydrodynamics and plasma physics~\cite{Benney-1967,Akhmanov-1968,Zakharov-1968,Taniuti-1968}. It is one of the central equations of nonlinear science, because it provides a unifying description of wave packets that travel in uniform media in the presence of dispersion and a weak nonlinearity. In this general frame, one considers  waves of the form
\beq
\psi(x,t)\exp i\left[\beta(\omega_0)x-\omega_0 t\right]+\cc,
\label{eq:generalwave}
\eeq
\textit{i.e.} an envelope $\psi$ that multiplies a sinusoidal carrier wave at a given frequency $\omega_0$. Above, ``$\cc$'' means ``complex conjugate'' and the function $\beta(\omega)$ contains all the relevant linear properties of the field in question. Provided that the envelope  varies slowly compared to the carrier, it generally satisfies
\beq
i\left(\pd\psi x+\beta'(\omega_0)\pd\psi t\right)-\frac{\beta''(\omega_0)}2\pd{^2\psi}{t^2}+\gamma\left|\psi\right|^2\psi=0,
\label{eq:NLS}
\eeq
which is the NLSE, where $\gamma$ is a nonlinear coefficient that depends on context. What makes the  success of Eq.~(\ref{eq:NLS}), besides its universality and robustness, is its relative simplicity and the fact that the envelope is decoupled from the underlying oscillations of the carrier wave. This decoupling appears quite reasonable owing to the separation of time scales between the two but the purpose of this communication is to challenge this view. It will be shown that the envelope can be pinned to the carrier wave in certain circumstances. 

As is well known, and as directly transpires from  Eq.~(\ref{eq:NLS}) the envelope travels at the group velocity $v_g=1/\beta'(\omega_0)$. On the other hand, the oscillations of the carrier wave move at the phase velocity  $v_p=\omega_0/\beta(\omega_0)$ in the limit of a vanishing nonlinearity ($\gamma=0$). From the point of view of the envelope, therefore, the carrier oscillations move at relative speed $v_p-v_g$ and whatever effect the latter could have on the former, it rapidly averages to zero over time. The possibility of a nontrivial interaction therefore rests on the assumption that 
\beq
\left|v_p-v_g\right|\ll v_p.
\label{eq:vp-vg}
\eeq
We thus specifically focus on wave systems whose linear response allows  that
\beq
\beta'(\omega_0)=\beta(\omega_0)/\omega_0
\label{vg=vp}
\eeq
for some $\omega_0$ and work in the vicinity of that angular frequency. A mathematically equivalent condition to (\ref{vg=vp}) is that the phase velocity passes by an extremum
\beq
v_p'(\omega_0)=0.
\label{eq:min_vp}
\eeq
Such a situation  can happen for instance with gravity-capillary waves, along elastic beams resting on a Winkler foundation or in cold plasmas~\cite{Taniuti-1968}. In optics and in acoustics, the constitutive properties of the medium rarely, if ever, allows  (\ref{eq:min_vp}) to happen in free space, but wave propagation in confined geometries leads to a greater variety of dispersion relations and makes that condition achievable. 

Balancing the last two terms of Eq.~(\ref{eq:NLS}), we see that $\psi$ evolves on a time scale $\left|\gamma/\beta''(\omega_0)\right|^{1/2}t$, to be compared with $\omega_0t$ for the carrier wave. Hence, the separation of timescales that underlies the validity of the NLSE  rests on the smallness of the following parameter
\beq
\veps = \left|\gamma/\beta''(\omega_0)\right|^{1/2}/\omega_0.
\label{def:vareps}
\eeq
Equivalently, $\veps$ measures the ratio of the wavelength of the sinusoidal oscillations to the width of the envelope. It is therefore essentially a geometrical parameter that can be identified independently of the physical context. Note that the smallness of $\veps$ does not imply the smallness of $\gamma$ in Eq.~(\ref{eq:NLS}), as it involves a ratio with $\omega_0$, which does not appear in that equation. Rather, $\veps\ll1$ is implicit in the derivation of Eq.~(\ref{eq:NLS}) from a given wave equation.

The NLSE only admits soliton solutions that travel at a constant speed. Within Eq.~(\ref{eq:NLS}), that speed can differ from $v_g(\omega_0)$, but in that case $\psi$ is modulated by a complex exponential, effectively shifting the frequency $\omega_0$ to a neighbouring value $\omega'$. As a result, the soliton speed predicted by Eq.~(\ref{eq:NLS}) remains in agreement with $v_g(\omega')$~\cite{Ablowitz-2011}.

What we will show is that, in a narrow but finite range of frequencies near $\omega_0$ defined by (\ref{eq:min_vp}), envelope solitons described to leading order by  Eq.~(\ref{eq:NLS})  do not move at a constant speed. Instead, their location is given by
\beq
x=v_p t +x_0(t),
\eeq
where
\beq
\ddot x_0(t)+\eta\frac{v_l^2}4\kappa\beta(\omega_0)   \sin\left[\kappa\beta(\omega_0)  x_0\right]=0,
\label{eq:central}
\eeq
\begin{align}
\kappa&=\left\{\begin{matrix}2, &\text{in the presence of inversion symmetry,}\\1, &\text{without inversion symmetry,}  \end{matrix}\right. &
\begin{matrix}\\\end{matrix}
 \end{align}
$\eta=\pm1$, and
\beq
v_l=K \veps^{-1/2}e^{-\kappa\pi /4\veps}v_p.
\label{eq:vl}
\eeq
Above, both $\eta$ and $K$ are model-dependent constants that  can only be obtained numerically.

Eq.~(\ref{eq:central}) is the central result of this paper. It holds both for bright and dark solitons. It can only be derived from the complete wave model, of which the NLSE is the leading order asymptotic reduction.  Hence, it partially invalidates the NLSE in the vicinity of an operating point given by Eq.~(\ref{eq:min_vp}). 
 
 The number $\kappa$ appearing in Eqs.~(\ref{eq:central}) and (\ref{eq:vl}) is connected to the set of harmonics generated by the nonlinearity. If the system is unchanged by a reversal of sign of the field, then the nonlinearity must be odd in that field. The lowest such nonlinearity is cubic and couples harmonics separated by $2\omega_0$ in the spectrum of the solution. Otherwise, the lowest possible nonlinearity is quadratic in the field. In the presence of the latter, successive peaks of the spectrum are separated by $\omega_0$. Hence the spacing between peaks is $\kappa\omega_0$ with $\kappa=1$ or 2, depending on the symmetry.

\begin{figure}
\hspace{-.3cm}
\includegraphics[width=8.cm]{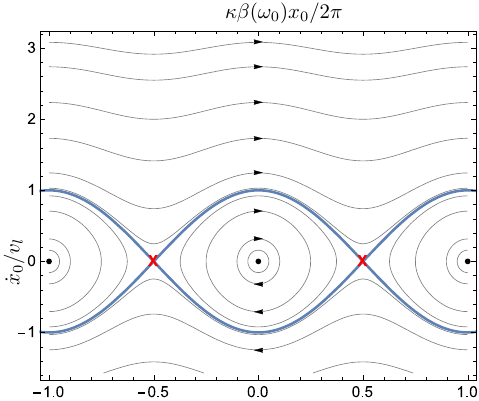}\\
\vspace{.12cm}
\hspace{-.17cm}\includegraphics[width=8cm]{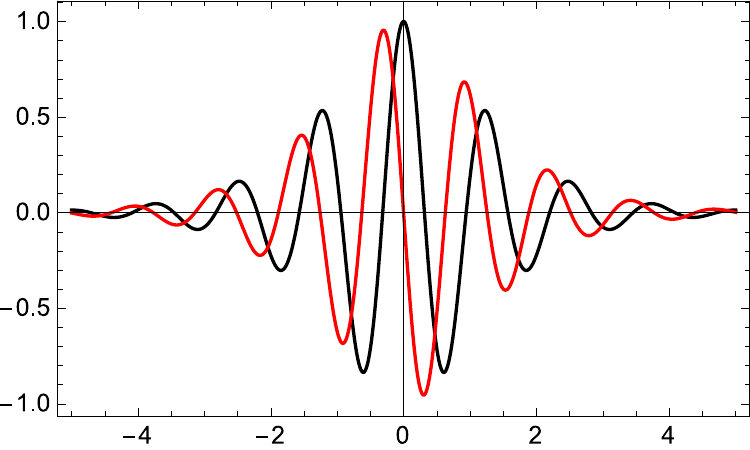}
\caption{Top: Phase portrait of Eq.~(\ref{eq:central}) in the case $\eta=1$. The blue lines are the separatrices that joins the unstable stationary points (red crosses). Black dots are stable stationary points. Note that well above the separatrices, $\dot x_0$ becomes nearly constant, so that the pinning force exerted by the carrier wave on the envelope becomes negligible. Bottom: wave packets travelling exactly at the phase velocity and corresponding to stable (black) and unstable (red) configurations in an inversion-symmetric system ($\kappa=2$). In this illustration, $\veps=0.2$.}\label{fig:pendulum}
\end{figure}

 According to Eq.~(\ref{eq:central}), the motion of the wave packet in a frame that moves at the phase velocity is analogous to that of a pendulum. It has stable stationary points given by $\kappa\beta(\omega_0)x_0=\left[2n+\left(\eta-1\right)/2\right]\pi$, where $n$ is an integer.  It also has unstable stationary points, $\kappa\beta(\omega_0)x_0=\left[2n+\left(\eta+1\right)/2\right]\pi$.    Those unstable points are connected by separatrices in the phase plane $(x_0,\dot x_0)$, see Fig.~\ref{fig:pendulum}. In the classical mechanical language, the closed phase plane trajectories inside the separatrices describe motion of libration; outside, the pendulum makes complete rotations about its point of attachment. The speed at $x_0=0$ that corresponds to the transition is $v_l$. In the phase portrait of Fig.~\ref{fig:pendulum}, we thus identify closed trajectories inside the separatrices as pertaining to the locking range, where the average speed $\left\langle v \right\rangle$ of the soliton is equal to $v_p$.
 
The pendulum equation (\ref{eq:central}) becomes compatible with the classical theory of soliton motion, Eq.~(\ref{eq:NLS}), when the kinetic energy of the pendular motion is very large. Indeed, if $|\dot x_0|\gg v_l$, then the phase portrait in Fig.~\ref{fig:pendulum} indicates that $\dot x_0$ becomes nearly constant. Hence, one recovers in that limit the continuous family of constant-speed solitons of the NLSE~\cite{Ablowitz-2011}. 

\begin{figure}
\includegraphics[width=7.8cm]{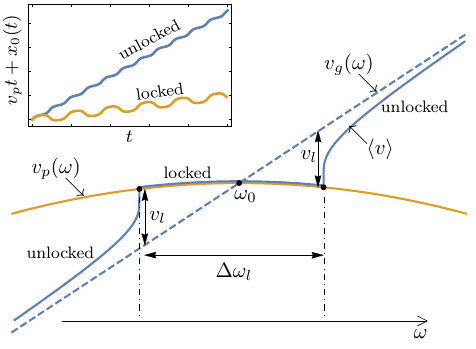}
\caption{Average soliton speed $\langle v\rangle$ in the vicinity of $\omega_0$. Inside the locking range $[\omega_0-\Delta\omega_l/2,\omega_0+\Delta\omega_l/2]$, the soliton envelope is locked to the phase velocity $v_p(\omega)$. Moving away from the locking range,  $\langle v\rangle$   gradually tends to the group velocity $v_g(\omega)$. Inset: examples of soliton trajectories just inside (orange) and just outside to the right of the locking range (blue).}
\label{fig:locking}
\end{figure}

\begin{figure}
\includegraphics[width=8cm]{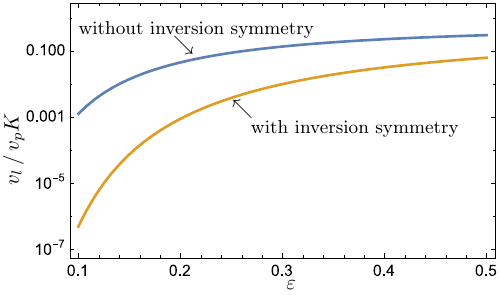}
\caption{Locking range, $\veps^{-1/2}\exp-\kappa\pi/4\veps$, as a function of the scale separation $\veps$ and of the system inversion symmetry ($\kappa=1,2$.)}
\label{fig:small}
\end{figure}

However, as the energy of the effective pendulum decreases, the phase portrait in Fig.~\ref{fig:pendulum} clearly indicates that the soliton velocity ceases to be constant and that the motion is unsteady. The integration of Eq.~(\ref{eq:central}) is a classical problem of mechanics. Outside the locking range, such that the pendulum makes complete rotations, the average speed  $\left\langle v \right\rangle$ of the soliton can be found as
\beq
\left\langle v \right\rangle 
= v_p(\omega)+\frac\pi2\left[\int_0^{\frac\pi2}\frac{\rd s}{\sqrt{\left(v_g(\omega)-v_p(\omega)\right)^2-v_l^2\cos s}}\right]^{-1}.
\eeq
This expression is illustrated in Fig.~\ref{fig:locking}. Within the locking range, the soliton oscillates about a coordinate that moves at the phase velocity. Quite remarkably, this oscillation is not due to noise, inhomogeneity of the supporting medium, or the presence of another soliton. It results solely from the oscillations of the carrier wave, which make an effective shallow periodic potential for the ``centre of mass'' of envelope. Fig.~\ref{fig:locking}  allows us to   ascertain the locking range in frequency as $2v_l/|v_g'(\omega_0)|$, that is
\beq
\Delta\omega_l =\frac{2v_l}{v_g^2(\omega_0)|\beta''(\omega_0)|}.
\eeq
The right hand side depends on the unknown constant $K$. Its dependence on $\veps$ is plotted in Fig.~\ref{fig:small} and shows that the effect is generally more pronounced for systems lacking inversion symmetry, due to the parameter $\kappa$. 

Also, $v_l$ makes more precise condition (\ref{eq:vp-vg}) on the closeness of the group and phase velocities. What is required in the present study is that
\beq
\left|v_p-v_g\right|=O\left(v_l\right)=O\left(\veps^{-1/2}e^{-\kappa\pi /4\veps}v_p\right).
\eeq

The quantity $v_l$ appearing in the pendulum equation is exponentially small, \textit{i.e.} smaller than any finite power of $\veps$. As a result, the physics presented here is not merely a high-order correction to the NLSE. In other word, it is of a different kind than the correction sometimes brought to the NLSE for very short pulses by adding third-order, fourth-order or higher order dispersion terms to the equation~\cite{Dudley-2006}. Rather, $v_l$ emerges because the multiple-scale expansion that underlies the NLSE actually generates a diverging series. As far back as 1857, George Gabriel Stokes studied diverging asymptotic series that approximate the Airy function~\cite{Stokes-1857}. He showed that the divergence is associated to the birth of exponentially small corrections in precise regions of the complex plane --corrections that grow exponentially as one moves away from their place of birth, to the point of completely invalidate the initial approximation. This process, known as Stokes phenomenon, also takes place in the present situation. We will show that a term of order $v_l^2$ will be switched on in the complex plane and will grow exponentially away from the centre of the wave packet, thereby threatening to invalidate the soliton approximation. To compensate for this correction, a second exponentially growing term is born that is proportional to $\ddot x_0$ so as to cancel the effect of the former. This summarises in a few words the technique of ``beyond-all-orders asymptotics", which has been applied to multiple-scale problem in only a few instances~\cite{YangAkylas-1997,Yang-1997,Wadee-1999,Wadee-2000,Calvo-2000,Wadee-2002,Adams-2003,Kozyreff-2006,Chapman-2009,Kozyreff-2009b,Dean-2011,Wadee-2012,Kozyreff-2013,Witt-2019}. Among these works, the one by  Yang and Akylas~\cite{YangAkylas-1997}  stands as particularly relevant. These authors studied the possibility of asymmetric gravity-capillary solitary waves in a simplified hydrodynamic model, namely the fifth order KdV equation. They were looking for constant wave profiles in a frame moving at $v_p$ and concluded that only symmetric profiles exist in that moving frame, \textit{i.e.}, that the wave either has a crest or a trough at its centre. This is consistent with the results quoted here: the nonlinearity in that example is quadratic ($\kappa=1$), so stationary solutions of the pendulum Eq.~(\ref{eq:central}) require either $x_0=0$ or $\beta(\omega_0)x_0=\pi$. In the follow-up research  coauthored by Calvo~\cite{Calvo-2000}, it was established that only one of them is stable, the one with a depression in the middle. This is again fully consistent with the present theory  with $\eta=-1$. Further, one can immediately read off from the pendulum equation that the rate of instability of the unstable stationary state is $0.5v_l\beta(\omega_0)$ and the scaling again agrees with Calvo \textit{et al.}. The present theory thus extends the pioneering work by these authors by placing it into a universal frame, completing the picture with dynamical asymmetrical profiles and including dark solitons. Note, finally, that a single computation can in principle allow one to determine the numerical constant $v_l$: fom what has just been said, it suffices to determine the rate of instability of the unstable stationnary profile.

Finally, it follows from what precedes that the carrier wave exerts a small periodic force on the center of mass of the envelope. This can of course alter the interaction between two solitons. At large distance,  two solitons are classically known to interact through a Toda potential, which may be repelling or attracting~\cite{YangBook-2010}. Such an interaction  potential can now contain ripples associated to the pendulum restoring force. As a result, new stable bound states can appear in the vicinity of the locking range, such that the two solitons, while interacting with each other, are at the same time tied to the carrier wave.

\section{A general wave model}
We will demonstrate our result on the following general class of wave equations
\beq
\pd{^2E}{x^2}+\int_{-\infty}^\infty\beta^2(\omega)\hat E(x,\omega)e^{-\ii\omega t}\rd \omega
= N\!\left[E,\eps\right]
\label{wave2}
\eeq
where 
\beq
\hat E(x,\omega)=\frac1{2\pi}\int_{-\infty}^\infty E(x,t)e^{\ii\omega t}\rd t
\eeq
is the Fourier transform of the scalar field $E(x,t)$ of interest and $N\![E,\eps]$ is the nonlinearity, which tends to zero as $\eps\to0$. We confine our study to a weak nonlinearity, \textit{i.e.} to $\eps\ll1$. In Appendix~\ref{appendix:eq1}, we show how the above equation naturally arises in electromagnetic and elastic and plasma wave propagation. The fifth order KdV equation, allowing waves only in one direction, is in fact not included in (\ref{wave2}). The agreement with the calculations of Calvo, Yang and Akylas on that equation support the claim that the results presented here are general and not tied to a particular wave model.  Notice that in the linear limit $\eps\to0$, Eq. (\ref{wave2})  is merely  a superposition of Helmholtz equations. 

The interest of formulating the wave equation in this slightly unusual way is that  it directly makes apparent the all-important dispersion function $\beta(\omega)$. If we momentarily neglect the right-hand side, it is immediate to see that the general wave solution moving in the positive $x$ direction is
\begin{align}
\eps&=0: &E(x,t)=\int_{-\infty}^\infty  A(\omega)e^{\ii\left[\beta(\omega)x-\omega t\right]}\rd\omega.
\end{align}
Hence $\beta(\omega)$ is indeed the wave number, or propagation constant, as a function of angular frequency. In what follows, we will adopt time and space units such that 
\begin{align}
\omega_0&=1, &\beta(\omega_0)&=1.
\end{align}
In these units, and in the absence of nonlinearity, the phase velocity is also unity at $\omega_0$.

When Eq.~(\ref{wave2}) models electromagnetic waves subjected to a Kerr nonlinearity, Maxwell's equations lead to
\beq
N\!\left[E,\eps\right]=  \frac{2}{3}\,\eps^2\pd{^2}{t^2}\left(E^3\right),
\label{wavenonlinearity0}
\eeq
where $\eps^2$ is a small parameter that is proportional to  the intensity of the wave and the factor 2/3 is introduced for later convenience. However, it is common practice to simplify the differential operator $\dd^2/\dd t^2$ above by $-\omega_0^2$. In order to simplify the algebra in this paper,  we will therefore present the calculation with the simpler cubic nonlinearity \beq
N\!\left[E,\eps\right]=- \frac{2}{3}\,\eps^2 E^3.
\label{wavenonlinearity}
\eeq
(Recall that $\omega_0=1$ in our choice of units.) We wish to stress, however, that we have also done the calculation with (\ref{wavenonlinearity0}) and obtained, as expected, the same result as with (\ref{wavenonlinearity}).  The algebra in the leading orders of the analysis is lightest with (\ref{wavenonlinearity}) but at later orders, the calculation becomes universal. We will therefore do the calculation explicitly with that nonlinearity and comment, whenever necessary in the course of our analysis, what happens when a quadratic, rather than a cubic nonlinearity is assumed.

Before embarking into a multiple-scale analysis, we note, as in \cite{Newell-book}, that for a wave packet with central frequency $\Omega$, we may expand $\beta(\omega)$ in Taylor series around $\Omega$ in the integral of Eq. (\ref{wave2}), giving
\begin{multline}
\int_{-\infty}^\infty\beta^2(\omega)\hat E(x,\omega)e^{-\ii\omega t}\rd \omega
=e^{-\ii\Omega t}\sum_n \frac1{n!}\left.\td{^n(\beta^2)}{\omega^n}\right|_{\Omega}\times\\
\int_{-\infty}^\infty\left(\omega-\Omega\right)^n\hat E(x,\omega)e^{-\ii(\omega-\Omega) t}\rd \omega+\cc\\
=e^{-\ii\Omega t}\sum_n \frac1{n!}\left.\pd{^n(\beta^2)}{\omega^n}\right|_{\Omega}\left(i\pd{}{t}\right)^n
\left(Ee^{\ii\Omega t}\right)+\cc\\
=e^{-\ii\Omega t}\beta^2\left(\Omega+\ii\dd/\dd t\right)\left(Ee^{\ii\Omega t}\right)+\cc.
\label{dispersion_operator}
\end{multline}

\paragraph*{\bf Remark:} The attentive reader will have noticed that we slightly changed the notation of the small parameter: $\eps$ vs $\veps$. The two differ by a numerical factor that we will specify shortly. Using $\eps$ will be more useful from a notational point of view in the asymptotic treatment to  follow, but $\veps$ is universally defined by (\ref{def:vareps}), independently of context.
\section{Multiple scales}
Let us construct a solution with the multiple-scale ansatz
\beq
E\sim \sum_{l\geq0}\eps^lE_l=\sum_{l\geq0}\sum_{m=-2l-1,}^{2l+1}\eps^l A_{l,m}\left(\xi,X\right)e^{\ii m(x-t)},
\label{expansion}
\eeq
with the constraint that
\beq
A_{l,-m}=\bar A_{l,m}
\eeq
on the real line for $E$ to be real (we use an overbar to denote complex conjugation.) Above, the function $A_{l,m}$ is the $O\left(\eps^l\right)$ contribution to the slow amplitude that modulates the $m^\text{th}$ harmonic of the fundamental carrier wave. Its evolution is assumed to take place on the following  spatio-temporal scales
\begin{align}
\xi&=\eps\left[t-\left(x-x_0\right)\left(1-\eps^2\nu \right)\right], &X=\eps^2 x,
\label{slowscales}
\end{align}
where $\eps^2\nu$ is a nonlinear correction to the phase velocity. Above, $\xi$ is a variable that is  attached to a frame moving at the speed $v_p\approx v_g$ and which serves to describe the envelope profile. On the other hand, $X$ is a slow evolution variable, akin to time, and which allows us to monitor slow changes of the envelope in the course of propagation.

Given (\ref{expansion}), we have, using (\ref{dispersion_operator}),
\begin{multline}
\int_{-\infty}^\infty\beta^2(\omega)\hat E(x,\omega)e^{-\ii\omega t}\rd \omega
=\\
\sum_{l, m}\eps^l e^{\ii m(x-t)} \left[\beta(m+\ii\eps \dd/\dd \xi)\right]^2A_{l,m}\left(\xi,X\right)\\
=\sum_{l,m,n}\eps^{l+n} e^{\ii m(x-t)}\Lambda^m_n \ii^n\pd{^nA_{l,m}}{\xi^n},
\end{multline}
where we have introduced
\beq
\Lambda^m_n =\frac1{n!}\left.\td{^n\beta^2(\omega)}{\omega^n}\right|_{m}.
\eeq
(Above, evaluation at $\omega=m$ means evaluation at $\omega=m\omega_0$ in a general set of units.)
Furthermore, with our multiple-scale ansatz, differentiation with respect to $x$ becomes
\beq
\pd{}{x}\to\pd{}{x}-\eps\left[1-\eps^2\nu\right]\pd{}{\xi}+\eps^2\pd{}{X}.
\eeq
As a result, $\dd^2/\dd x^2$ becomes
\begin{multline}
\pd{^2}{x^2}\to\pd{^2}{x^2}-2\eps\pd{^2}{x\dd\xi}+\eps^2\left(\pd{^2}{\xi^2}+2\pd{^2}{x\dd X}\right)\\
+2\eps^3\left(\nu\pd{^2}{x\dd \xi}-\pd{^2}{\xi \dd X}\right)+\eps^4\left(\pd{^2}{X^2}-2\nu\pd{^2}{\xi^2}\right)\\
+2\eps^5\nu\pd{^2}{\xi\dd X}+\eps^6\nu^2\pd{^2}{\xi^2}.
\end{multline}
Above the $O\left(\eps^5\right)$ terms are unimportant for the leading orders of the asymptotic analysis and are also negligible in first approximation when it comes to late-terms of the expansion. However, we give them for the sake of completeness. We thus have
\begin{multline}
\pd{^2E}{x^2}\sim
\sum_{lm}\eps^l e^{\ii m(x-t)}\left(
-m^2A_{l,m}-2\ii m \pd{A_{l-1,m}}{\xi}+ \right.\\
\pd{^2A_{l-2,m}}{\xi^2}+2\ii m\pd{A_{l-2,m}}X
+2im\nu  \pd{A_{l-3,m}}{\xi}
-2 \pd{^2A_{l-3,m}}{\xi \dd X}\\
+\pd{^2A_{l-4,m}}{X^2}-2\nu \pd{^2A_{l-4,m}}{\xi^2}
+2\nu\pd{^2A_{l-5,m}}{\xi\dd X}
\left.
+\nu^2\pd{^2A_{l-6,m}}{\xi^2}
\right)
.
\label{d2E/dx2}
\end{multline}
Finally,
\beq
E^3=
\sum_{ll'l"mm'm"}\eps^l
A_{l',m'}A_{l",m"}A_{l-l'-l",m-m'-m"}e^{\ii m(x-t)}.
\eeq
Putting everything together, we have to solve, for each order $l\geq0$ and each  harmonic $m$,
\begin{multline}
-m^2A_{l,m}-2\ii m  \pd{A_{l-1,m}}{\xi}
+ \pd{^2A_{l-2,m}}{\xi^2}
+2\ii m\pd{A_{l-2,m}}X\\
+2im\nu  \pd{A_{l-3,m}}{\xi}-2 \pd{^2A_{l-3,m}}{\xi \dd X}
+\pd{^2A_{l-4,m}}{X^2}
-2\nu \pd{^2A_{l-4,m}}{\xi^2}\\
+2\nu\pd{^2A_{l-5,m}}{\xi\dd X}
+\nu^2\pd{^2A_{l-6,m}}{\xi^2}
+\sum_n\Lambda^m_n \ii^n\pd{^nA_{l-n,m}}{\xi^n}\\
=-\frac23\sum_{l'l"m'm"} A_{l',m'}A_{l"m"}A_{l-2-l'-l",m-m'-m"} .
 \label{recur1}
\end{multline}
This is the general multiple-scale translation of Eq.~(\ref{wave2}) subject to the nonlinearity (\ref{wavenonlinearity}). In what follows, we will first solve the above equations up to $l=2$ to derive the NLSE for the amplitude $A_{0,1}$. We will next investigate the recurrence for $l\gg1$ and show that $A_{lm}$ grows factorially with $l$ in that limit, making the asymptotic series (\ref{expansion}) diverge. Finding the precise way in which this divergence occurs, we will be able to truncate  (\ref{expansion}) optimally as
\beq
E=\sum_{l=0}^{L-1}\eps^lE_l+R
\label{expansion2}
\eeq
for some large $L$ and derive an equation for the remainder $R$. The law governing the locking of the envelope to the carrier wave is contained in $R$.

\subsection{Leading orders}
At $l=0,m=\pm1$, we obtain
\beq
\left(\beta(1)^2-1\right)A_{0,1}=0,
\eeq
which is automatically satisfied, since $\beta(1)=1$.
At $l=1,m=1$, we have
\beq
\ii\left(\Lambda^1_1-2\right)\pd{A_{0,1}}\xi =2\ii\frac{v_p-v_g}{v_g}\pd{A_{0,1}}\xi
=0.
\label{vp/vg-1}
\eeq
By assumption, the difference $v_p-v_g$  is exponentially small so that the above equation is automatically satisfied at $O(\eps)$ of our calculation.

Next, the equation for $l=1,m=3$, yields
\begin{align}
\left(\beta(3)^2-9\right)A_{1,3}&=0, &\to A_{1,3}&=0,
\end{align}
$\beta(3)^2$ being generally different from $9$, due to dispersion. Finally, the $l=2,m=1$ equation is
\beq
\pd{^2A_{0,1}}{\xi^2}+2\ii\pd{A_{0,1}}{X}-\Lambda^1_2\pd{^2A_{0,1}}{\xi^2}=-2A_{0,1}^2A_{0,-1},
\eeq
or, equivalently
\beq
\ii\pd{A_{0,1}}{X}+\frac{1-\Lambda^1_2}{2}\pd{^2A_{0,1}}{\xi^2}+\left|A_{0,1}\right|^2A_{0,1}=0.
\eeq
Evaluating $\Lambda^1_2$, we get
\beq
\Lambda^1_2=\beta'(1)^2+\beta(1)\beta''(1)=1+\beta_2,
\eeq
where $\beta_2=\beta''(1)$. Eventually, we find the classical NLSE:
\beq
\ii\pd{A_{0,1}}{X}-\frac{\beta_2}{2}\pd{^2A_{0,1}}{\xi^2}+\left|A_{0,1}\right|^2A_{0,1}=0.
\label{NLS}
\eeq
One may check that the above eqaution is indeed equivalent to  Eq.~(\ref{eq:NLS}) by substituting $\psi=A_{0,1}(\eps(t-\beta'(\omega_0)x),\eps^2x)$ in the latter and setting $\gamma=\eps^2$.

\subsection{Soliton solution}\label{sec:solitonsol}

Given the sign of the nonlinear term in (\ref{NLS}), the bright soliton solution is obtained when $\beta_2<0$ and is given by
\beq
A_{0,1}=e^{\ii X/2} \sech\left(\frac{\xi}{|\beta_2|^{1/2}}\right),
\eeq
while the leading order amplitude of the $m=-1$ harmonic is given by
\beq
A_{0,-1}=e^{-\ii X/2} \sech\left(\frac{\xi}{|\beta_2|^{1/2}}\right).
\eeq
On the other hand, if  $\beta_2>0$, the NLSE admits the dark soliton solution
\beq
A_{0,\pm1}=e^{\pm\ii X} \tanh\left(\frac{\xi}{|\beta_2|^{1/2}}\right).
\eeq
Hence, the leading-order soliton solution of the complete model is
\beq
E_0=\left\{
\begin{matrix}
2\sech\left(\frac{\xi}{|\beta_2|^{1/2}}\right)\cos(x-t+X/2)&\text{  if \quad}\beta_2<0,\\
2\tanh\left(\frac{\xi}{|\beta_2|^{1/2}}\right)\cos(x-t+X)&\text{  if \quad}\beta_2>0.
\end{matrix}
\right.
\label{leadingsol}
\eeq
Importantly for what follows, both solutions above have complex singularities at $\xi=\ii(n+1/2)\pi |\beta_2|^{1/2}$. In particular, for the bright soliton solution
\beq
A_{0,\pm1}\sim \frac{\ii |\beta_2|^{1/2} e^{\pm\ii X/2}}{\xi_0-\xi}
\eeq
as $\xi\to  \xi_0=\ii\pi |\beta_2|^{1/2}/2$ while, in the same limit, the dark soliton solution diverges as
\beq
A_{0,\pm1}\sim  \frac{-\, \beta_2^{1/2} e^{\pm\ii X}}{\xi_0-\xi}.
\eeq
We may summarise the two possible behaviours in the vicinity of $\xi_0$ as
\beq
A_{0,\pm1}\sim \frac{i^* |\beta_2|^{1/2} e^{\pm \ii X/2^*}}{\xi_0-\xi}.
\eeq
where we introduce the notation
\begin{align}
i^*&=\ii, &2^*&=2, &\text{  if \quad}\beta_2<0,\\
i^*&=-1, &2^*&=1, &\text{  if \quad}\beta_2>0.
\end{align}
Note that the star does not mean complex conjugation -we use an overbar for that purpose. Next,   recalling that $X=\eps^2x$ in (\ref{leadingsol}), the nonlinear correction to the phase velocity is
\beq
\eps^2\nu
=\left\{
\begin{matrix}
-\eps^2/2&\text{  if \quad}\beta_2<0,\\
-\eps^2&\text{  if \quad}\beta_2>0,
\end{matrix}
\right.
\eeq
which can be summarised by writing $\nu=-1/2^*$ in the starred notation.
Finally, comparing the scales of the envelope and the carrier wave in (\ref{leadingsol}), we see that the small parameter $\veps$ of the introduction is related to $\eps$ as
\beq
\veps=\eps/|\beta_2|^{1/2}.
\eeq

\subsection{Galilean invariance}
The above formulas for $A_{0,1}$, whether they describe bright or dark solitons, belong to a family of solutions of the NLSE obtained by the transformation rule
\begin{align}
\xi&\to\xi+cX,
&
 i X/2^* &\to   i \left(X/2^*+\frac{c^2X}{2\beta_2}+\frac{c\xi}{\beta_2}\right),
\end{align}
where $c$ is a constant parameter that produces a change $\eps c$ of the envelope velocity. The second transformation rule above ensures that this change of group velocity is accompanied by a change of frequency that is compatible with the dispersion relation $\beta(\omega)$.

\subsection{Slowly accelerating soliton}
If we now allow $c$ to vary slowly in the course of propagation, then Galilean invariance is broken. Letting $c'(X)\ll1$, the bright soliton solution is changed, to first order, as
\begin{equation}
A_{0,1}
\sim\left[\sech\left(\zeta\right)-\td{c}{X}\left( R_a+iX\zeta\right)\right] e^{\ii \left[X/2+\left(c^2X/2+c\xi\right)/\beta_2\right]},
\label{accelerate1}
\end{equation}
where
\beq
\zeta = \frac{\xi+c X}{|\beta_2|^{1/2}},
\eeq
and  $R_a$ is the correction due to the acceleration:
\begin{equation}
R_a= \frac{1+e^{4\zeta}-2\zeta+2\zeta^2-2e^{2\zeta}\left(5-5\zeta+\zeta^2\right)}{2\left(e^{3\zeta}+2e^{\zeta}+e^{-\zeta}\right)}.
\end{equation}
Note that the far-field behaviour of $R_a$ is
\begin{align}
R_a&\sim\frac{e^\zeta}{2}, &\zeta&\to\infty,
\label{farfieldRabright}
\end{align}
which, in principle, renders (\ref{accelerate1}) physically unacceptable. However, another exponentially growing term is hidden in the remainder $R$ of (\ref{expansion2}), which can compensate $R_a$ and allow the existence of accelerating solitons.

For dark solitons ($\beta_2>0$)  the same thing happens:
\begin{equation}
A_{0,1}
\sim\left[\tanh\left(\zeta\right)+\td{c}{X}\left( R_a+iX\zeta\right)\right] e^{\ii \left[X+\left(c^2X/2+c\xi\right)/\beta_2\right]},
\label{accelerate2}
\end{equation}
this time with
\begin{multline}
R_a=\frac{\sech^2\zeta}{32}\left[-7+12\zeta-8\zeta^2+8\cosh2\zeta\right.\\
\left.+8(1-\zeta)\sinh2\zeta+\cosh4\zeta+\sinh4\zeta\right]
\end{multline}
and
\begin{align}
R_a&\sim\frac{e^{2\zeta}}{8}, &\zeta&\to\infty.
\label{farfieldRadark}
\end{align}

\section{Late-term expansion}
We now investigate the large-$l$ behaviour of the amplitudes $A_{lm}$. We first note that the nonlinearity (\ref{wavenonlinearity}) only couple odd harmonics of the fundamental one and that two new harmonics appear at each order of the calculation. As a result non-zero amplitudes $A_{lm}$ only exist for
\begin{align}
m \text{ odd},&& |m|\leq 2\lfloor l/2\rfloor +1.
\end{align}
[\textit{With a quadratic nonlinearity, even harmonics $m$ exist too.}]
As $l$ becomes large, we anticipate from similar calculations~\cite{Kozyreff-2006,Chapman-2009,Kozyreff-2009b,Dean-2011,Kozyreff-2013} that these amplitude grow in size as $\Gamma(l+\alpha_m)$ for some $\alpha_m$. In order to determine this number, it is sufficient to study the system (\ref{recur1}) in the vicinity of the singularity $\xi_0$. This is what we do first.

\subsection{Inner expansion near $\xi=\xi_0$}
In the vicinity of $\xi_0$, we assume that
\beq
A_{l,m}\sim \frac{\Gamma\left(l+\alpha_m\right) B_{l,m}e^{\ii mX/2^*}}{\left(\xi_0-\xi\right)^{l+1}},
\label{inneransatz}
\eeq
where $\alpha_m$ is a constant to be determined. Substituting in Eq.~(\ref{recur1}) and factoring out the gamma function, we obtain, as $\xi\to\xi_0$, 
\begin{multline}
-m^2B_{l,m}-2\ii m   \frac{l \Gamma(l+\alpha_m-1)}{\Gamma(l+\alpha_m)} B_{l-1,m}\\
+  \frac{l (l-1)\Gamma(l+\alpha_m-2)}{\Gamma(l+\alpha_m)} B_{l-2,m}\\
+\sum_n\Lambda^m_n \ii^n \frac{l!\Gamma(l+\alpha_m-n)}{(l-n)!\Gamma(l+\alpha_m)}B_{l-n,m}\\
=-\frac23\sum_{l'l"m'm"}\frac{\Gamma(l'+\alpha_{m'})\Gamma(l''+\alpha_{m''})}{\Gamma(l+\alpha_m)}B_{l',m'}B_{l",m"}\\ \times
\Gamma(l-2-l'-l"+\alpha_{m-m'-m"})B_{l-2-l'-l",m-m'-m"}.
 \label{recur2}
\end{multline}
Compared to (\ref{recur1}), Eq.~(\ref{recur2}) is a set of algebraic equations, rather than differential ones, and are therefore more tractable. They should in principle be solved by recurrence, starting with
\beq
B_{0,\pm1}=i^*|\beta_2|^{1/2}/\Gamma(\alpha_1).
\label{inicon}
\eeq
In the large-$l$ limit, the equations simplifies to
\begin{multline}
-m^2B_{l,m}-2\ii m   B_{l-1,m}
+  B_{l-2,m}\\
+\sum_n\Lambda^m_n \ii^n B_{l-n,m}
=O(1/l).
 \label{recur3}
\end{multline}
We may then look for a solution of the form
\beq
B_{l,m}\sim (\ii/\kappa)^{l} b_m^{(0)}+O(1/l),
\label{Blm1}
\eeq
where $\kappa$ is to be determined. This yields
\beq
\left(\sum_n\Lambda^m_n \kappa^n-m^2-2 m \kappa  -\kappa^2
\right) b_m^{(0)}
=0.
 \label{recur4}
\eeq
Recalling the definition of the coefficients $\Lambda^m_n$, the above equation can be written more simply as
\beq
\left[\beta^2\left(m+\kappa\right)- \left(m+ \kappa\right)^2\right]b_m^{(0)}=0.
\eeq
Given that $\beta(\pm1)^2=1$, we thus have
\beq
m+\kappa = \pm1,
\label{kappa}
\eeq
which can be solved for $\kappa$ as a function of the harmonic $m$. Since $m$ is odd, the possible  values of $\kappa$ are
\beq
\kappa=\pm2,\pm4,\pm6,\ldots.
\label{kappa2}
\eeq
In particular,  $\kappa=2$ is obtained for $m=-1,-3$, while $\kappa=-2$ corresponds to $m=1,3$. [\textit{With a quadratic nonlinearity, even values are allowed for $m$, so we have instead $\kappa=\pm1,\pm2,\pm3,\ldots$. In particular, $\kappa=1$ is obtained for $m=-2,0$, while $\kappa=-1$ corresponds to $m=0,2$.}]

Given the $\kappa^{-l}$ dependance in Eq.~(\ref{Blm1}), we may expect that the Fourier modes corresponding to smallest absolute value of $\kappa$ are those that will matter at very large orders and that the other Fourier components will be subdominant as $l\to\infty$.

To make further progress, let us generalise Eq.~(\ref{Blm1}) as
\beq
B_{l,m}\sim (\ii/\kappa)^{l} \left(b_m^{(0)}+l^{-1}b_m^{(1)}+l^{-2}b_m^{(2)}+\dotsb\right),
\label{Blm2}.
\eeq
Then we find (see Appendix~\ref{appendix:gq(l)}) that
\beq
\frac{\ii^n l!\Gamma(l+\alpha_m-n)}{(l-n)!\Gamma(l+\alpha_m)}B_{l-n,m}
\sim
(\ii/\kappa)^{l}\sum_{q=0}^n \frac{n! \kappa^n}{(n-q)!} g_q(l)
\label{factor}
\eeq
with
\begin{align}
g_0(l)&\sim b_m^{(0)}+l^{-1}b_m^{(1)}+l^{-2}b_m^{(2)}+\dotsb,\label{g0}\\
g_1(l)&\sim\frac{(1-\alpha_m)b_m^{(0)}}{l}+ \frac{(2-\alpha_m)b_m^{(1)}+(1-\alpha_m)^2 b_m^{(0)}}{l^2},\label{g1}\\
g_2(l)&\sim\frac{(1-\alpha_m)(2-\alpha_m)b_m^{(0)}}{2l^2},\label{g2}\\
\vdots \non\\
g_q(l)&= O(1/l^q).
\end{align}
Now notice that
\begin{align}
\sum_n \Lambda^m_n\kappa^n&=\beta^2(m+\kappa),\\
\sum_n n \Lambda^m_n\kappa^{n}&=\kappa\td{}{\kappa}\left(\beta^2(m+\kappa)\right),\\
\sum_n n(n-1) \Lambda^m_n\kappa^{n}&=\kappa^2\td{^2}{\kappa^2}\left(\beta^2(m+\kappa)\right),\ldots
\end{align}
Hence, for any $q$,
\beq
\sum_n\Lambda^m_n \frac{n!\kappa^{n}}{(n-q)!} =\kappa^q\td{^q}{\kappa^q}\left(\beta^2(m+\kappa)\right),
\eeq
Combining this last expression with Eq.~(\ref{factor}), we find that the various terms in the l.h.s. of Eq.~(\ref{recur2}), can be written as
\begin{align}
\frac{\ii  l \Gamma(l+\alpha_m-1)}{\Gamma(l+\alpha_m)} B_{l-1,m}\sim& (\ii/\kappa)^l [g_0(l)+g_1(l)]\kappa,\\
\frac{\ii^2 l (l-1)\Gamma(l+\alpha_m-2)}{\Gamma(l+\alpha_m)} B_{l-2,m}
\sim&(\ii/\kappa)^l [g_0(l)+2g_1(l)\non\\
&+2g_2(l) ]\kappa^2.
\end{align}
and
\begin{multline}
\sum_n\Lambda^m_n \ii^n \frac{l!\Gamma(l+\alpha_m-n)}{(l-n)!\Gamma(l+\alpha_m)}B_{l-n,m}
=\\
(\ii/\kappa)^{l} \bigg[
g_0(l)\beta^2(m+\kappa)
+g_1(l)\kappa \td{}{\kappa}\left(\beta^2(m+\kappa)\right) \\
+g_2(l)\kappa^2 \td{^2}{\kappa^2}\left(\beta^2(m+\kappa)\right)+\dotsb\bigg].
\end{multline}
Putting everything together, the l.h.s. of (\ref{recur2}) is
\begin{multline}
(\ii/\kappa)^l\bigg\{
g_0(l)  \left[\beta^2\left(m+\kappa\right)- \left(m+\kappa\right)^2\right]\\
+g_1(l) \kappa\left[\td{}{\kappa}\beta^2\left(m+\kappa\right)- 2(\kappa+m) \right]\\
+g_2(l)\kappa^2\left[\td{^2}{\kappa^2}\beta^2\left(m+\kappa\right)-2\right]+O(1/l^3)\bigg\}.
\label{lhs2}
\end{multline}
Above, we compute
\begin{align}
\td{\beta^2\left(\omega\right)}{\omega}&=2\beta\left(\omega\right) \beta'\left(\omega\right), \\
\td{^2\beta^2\left(\omega\right)}{\omega^2}&=2 \beta'\left(\omega\right) ^2 
+2 \beta\left(\omega\right)  \beta''\left(\omega\right)
\end{align}
If $m+\kappa=1$, we find that $\left(\rd/\rd \kappa\right)\beta^2\left(m+\kappa\right)=2$ and $\left(\rd^2/\rd \kappa^2\right)\beta^2\left(m+\kappa\right)=2+2\beta_2$. On the other hand, from the fact that $\beta^2(-\omega)=\beta^2(\omega)$, we deduce that if $m+\kappa=-1$, then $\left(\rd/\rd \kappa\right)\beta^2\left(m+\kappa\right)=-2$ and $\left(\rd^2/\rd \kappa^2\right)\beta^2\left(m+\kappa\right)=2+2\beta_2$. In both cases, the first two terms in the curly bracket of (\ref{lhs2}) vanish. Hence, the l.h.s. of Eq.~(\ref{recur2}) is found to be asymptotic to
$
-2\beta_2(\ii/\kappa)^{l-2} g_2(l)
$, 
\textit{i.e.} to
\beq
-\beta_2(\ii/\kappa)^{l-2} (1-\alpha_m)(2-\alpha_m)b_m^{(0)} /l^2
\label{lhs}
\eeq
Therefore, the right hand side of (\ref{recur2}) must be $O(l^{-2})$. The right hand side mixes and couples different harmonics. It is natural to consider a couple of harmonics related by the same $\kappa$, as in Eq.~(\ref{kappa}). Let them be $m$ and $m+2$ and let us assume that $\alpha_m=\alpha_{m+2}=\alpha$. It is easy to see, that in the large-$l$ limit, the leading terms in the right hand side of (\ref{recur2}) will be those for which two of the three indices $l',l'',l-2-l'-l''$ are zero. Thus, using (\ref{inicon}), (\ref{kappa}) and the definition of $i^*$, the right hand side of (\ref{recur2}) is asymptotic to
\begin{multline}
\frac{-2(\ii/\kappa)^{l-2}}{l^2}\sum_{m',m''}
\Gamma(\alpha_{m'})\Gamma(\alpha_{m''}) B_{0,m'}B_{0,m''}
b^{(0)}_{m-m'-m''}\\
=\frac{-2(\ii/\kappa)^{l-2}}{l^2}\left(i^{*}\right)^2|\beta_2| \sum_{m',m''} b^{(0)}_{m-m'-m''}\\
=\frac{-2\beta_2(\ii/\kappa)^{l-2}}{l^2} \sum_{m',m''} b^{(0)}_{m-m'-m''}\\
=\frac{-2\beta_2(\ii/\kappa)^{l-2}}{l^2}\left(b^{(0)}_{m-2}+2b^{(0)}_{m}+b^{(0)}_{m+2}\right)
\end{multline}
Eventually Eq.~(\ref{recur2}) evaluated for the two values, $m$ and $m+2$, corresponding to a given $\kappa$, yield
\begin{align}
\left[\left(1-\alpha\right)\left(2-\alpha\right)-4\right]b_m^{(0)}-2b_{m+2}^{(0)}&=0,\\
\left[\left(1-\alpha\right)\left(2-\alpha\right)-4\right]b_{m+2}^{(0)}-2b_{m}^{(0)}&=0.
\end{align}
Then we find that either $b_{m+2}^{(0)}=b_m^{(0)}$ and
\begin{align}
(1-\alpha)(2-\alpha)-6&=0, 
&\to \alpha\in\{-1,4\}.
\end{align}
 or $b_{m+2}^{(0)}=-b_{m}^{(0)}$, in which case
\begin{align}
(1-\alpha)(2-\alpha)-2&=0,
&\to \alpha\in\{0,3\}.
\end{align}
Out of all the above possibilities, the case where $\alpha=4$ will yield the dominant  factorial growth as $l\gg1$. We therefore only keep that value into consideration in what follows. We have thus shown that
\begin{equation}
E_{l}\sim i^*\sum_\kappa\sum_{m=-\kappa\pm1} \lambda_\kappa \frac{\left(\ii/\kappa\right)^l\Gamma(l+4)} {\left(\xi_0-\xi\right)^{l+1}} e^{\ii m(x-t+X/2^*)}
\label{innersol1}
\end{equation}
at order $l$ in the expansion of $E$. Above, $\lambda_\kappa$ is a constant that can only be computed numerically or by actually solving the recurrence equations (\ref{recur2}) all the way from $l=0$ to $l\gg1$. Note that there is a similar expression near $\xi=-\xi_0$ and indeed in the vicinity of all complex singularities of the leading order soliton. Although the constant $\lambda_\kappa$ is not universal, we note that, after making the change $B_{l,m}=(\ii/\kappa)^l b_{l,m}$, the resulting equations for $b_{l,m}$ have only real coefficients. Hence, given the factor $i^*$ in (\ref{innersol1}), we may deduce that $\lambda_\kappa$ is real.

\subsection{Outer expansion away from $\xi=\xi_0$}

The result of our investigation in the vicinity of $\xi_0$, Eq.~(\ref{innersol1}), suggests  to look for contributions in the late-term expansion of the form
\begin{align}
E_{l}\sim \sum_\kappa\sum_{m=-\kappa\pm1} \lambda_\kappa \frac{\left(\ii/\kappa\right)^l\Gamma(l+4)}{\left(\xi_0-\xi\right)^{l+4}} f_m(\xi,X)e^{\ii m (x-t)}.
\end{align}
Note that the exponent in the denominator was slightly changed with respect to that in (\ref{innersol1}). This small variation, which is allowed by the undetermined factor $f_m(\xi,X)$, will significantly simplify the ensuing algebra.
Thus, the $A_{lm}$ are of the form
\beq
A_{l,m}=\lambda_\kappa \frac{\left(\ii/\kappa\right)^l\Gamma(l+4)}{\left(\xi_0-\xi\right)^{l+4}} f_m(\xi,X)
\label{outeransatz}
\eeq
Considering (\ref{recur1}), we see that we will need to evaluate expressions of the form
\begin{multline}
\pd{^nA_{l-n,m}}{\xi^n}
=
\lambda_\kappa \left(\ii/\kappa\right)^{(l-n)}\Gamma(l+4-n)\\
\times\pd{^n}{\xi^n}\frac{f_m(\xi,X)}{\left(\xi_0-\xi\right)^{l+4-n}}  
=\lambda_\kappa \left(\ii/\kappa\right)^{(l-n)}\Gamma(l+4-n)  \\
\times \sum_{o=0}^n
 \begin{pmatrix}n\\o\end{pmatrix}
\frac{\Gamma(l+4-o)}{\Gamma(l+4-n)}\frac{1}{\left(\xi_0-\xi\right)^{l+4-o}} \pd{^o f_m(\xi,X)}{\xi^o} \\
=
\lambda_\kappa \left(\ii/\kappa\right)^{(l-n)}  
 \sum_{o=0}^n
 \begin{pmatrix}n\\o\end{pmatrix}
 \frac{\Gamma(l+4-o)}{\left(\xi_0-\xi\right)^{l+4-o}}\pd{^o f_m(\xi,X)}{\xi^o} ,
\end{multline}
where we have used the binomial formula for the derivative of a product. Developing this expression further:
\begin{align}
&\pd{^nA_{l-n,m}}{\xi^n}
=
\lambda_\kappa \frac{\left(\ii/\kappa\right)^{l}\Gamma(l+4)}{\left(\xi_0-\xi\right)^{l+4}}  \non\\
&\times\left(\kappa/i\right)^n \sum_{o=0}^n
 \begin{pmatrix}n\\o\end{pmatrix}
 \frac{\Gamma(l+4-o)}{\Gamma(l+4)} \left(\xi_0-\xi\right)^{o}\pd{^o f_m}{\xi^o} .
\end{align}
Expanding the sum above up to order $1/l^2$:
\begin{align}
&\sum_{o=0}^n
 \begin{pmatrix}n\\o\end{pmatrix}
 \frac{\Gamma(l+4-o)}{\Gamma(l+4)} \left(\xi_0-\xi\right)^{o}\pd{^o f_m}{\xi^o} \non\\
 &=f_m+\frac{n}{l+3}\left(\xi_0-\xi\right) \pd{f_m}{\xi} \non\\
& +\frac{n(n-1)}{2(l+3)(l+2)}\left(\xi_0-\xi\right)^2 \pd{^2f_m}{\xi^2} +\ldots\non\\
 &\sim
  f_m+n\left(\frac1l-\frac3{l^2}\right)\left(\xi_0-\xi\right) \pd{f_m}{\xi} \non\\
 &+\frac{n(n-1)}{2l^2}\left(\xi_0-\xi\right)^2 \pd{^2f_m}{\xi^2} +O\left(l^{-3}\right)
\end{align}
Hence, following a similar path as in the previous section, the dispersive term in the equation for $A_{lm}$ is
\begin{align}
&\sum_n\Lambda^m_n \ii^n\pd{^nA_{l-n,m}}{\xi^n}\non\\
&\sim
\lambda_\kappa \frac{\left(\ii/\kappa\right)^{l}\Gamma(l+4)}{\left(\xi_0-\xi\right)^{l+4}} 
\bigg(
  f_m\beta^2(m+\kappa)+\non\\
 & \left(\frac1l-\frac3{l^2}\right)\left(\xi_0-\xi\right) \pd{f_m}{\xi} \kappa\td{}{\kappa}\left(\beta^2(m+\kappa)\right)\non\\
 &+\frac{1}{2l^2}\left(\xi_0-\xi\right)^2 \pd{^2f_m}{\xi^2} 
\kappa^2\td{^2}{\kappa^2}\left(\beta^2(m+\kappa)\right)
 +O\left(l^{-3}\right)
\bigg)\non\\
&=
\lambda_\kappa \frac{\left(\ii/\kappa\right)^{l}\Gamma(l+4)}{\left(\xi_0-\xi\right)^{l+4}} 
\bigg(
  f_m +\non\\
&2\kappa (m+\kappa)  \left(\frac1l-\frac3{l^2}\right)\left(\xi_0-\xi\right) \pd{f_m}{\xi} \non\\
 &+\frac{\kappa^2(1+\beta_2)}{l^2}\left(\xi_0-\xi\right)^2 \pd{^2f_m}{\xi^2} 
 +O\left(l^{-3}\right)
\bigg),
\end{align}
where we recall that $m+\kappa=\pm1$. Eventually we find that  the l.h.s. of (\ref{recur1}) is
\begin{multline}
\lambda_\kappa \frac{\left(\ii/\kappa\right)^{l}\Gamma(l+4)}{\left(\xi_0-\xi\right)^{l+4}}    \frac{\kappa^2\left(\xi_0-\xi\right)^2}{l^2}\\
 \times
  \left[ 
2 (m+\kappa)\left(-\nu\kappa f_m-\ii \pd{f_m}{X}\right)+ \beta_2  \pd{^2f_m}{\xi^2}
+O\left(l^{-3}\right)	 \right]
\label{lhs4}	
\end{multline}
As for the nonlinear terms of Eq.~(\ref{recur1}), the leading order terms in the large-$l$ limit are 
\begin{multline}
-2\sum_{m'm"}A_{0,m'}A_{0,m"}2A_{l-2,m-m'-m"}\\
 \sim 2 \lambda_\kappa \frac{\left(\ii/\kappa\right)^{l}\Gamma(l+4)}{\left(\xi_0-\xi\right)^{l+4}}\frac{\kappa^2\left(\xi_0-\xi\right)^2}{l^2}
 \\  \times
 \left(A_{0,-1}^2f_{m+2}+2A_{0,-1}A_{0,1}f_m+A_{0,1}^2f_{m-2}\right)
 \label{rhs4}
\end{multline}
Eventually, we obtain
\begin{multline}
\left(m+\kappa\right)\left(\nu\kappa\beta_1f_m+\ii \pd{f_m}{X}\right)- \frac{\beta_2}2  \pd{^2f_m}{\xi^2}\\
+A_{0,1}^2f_{m-2}+2A_{0,-1}A_{0,1}f_m+A_{0,-1}^2f_{m+2}=0,
\label{eqforfm}
\end{multline}
Let now $f_m$ be given by
\begin{align}
f_m&=F(\xi)e^{\ii m X/2^*},
&m=-\kappa\pm1
\end{align}
and be zero otherwise. Then,
\beq
\left(m+\kappa\right)\left(\nu\kappa-\frac{m}{2^*}\right)F- \frac{\beta_2}2  \td{^2F}{\xi^2}
+3|A_{0,1}|^2F=0.
\eeq
The equation for $F$ is identical for the two possible values of $m$ because $\nu=-1/2^*$, so that
\beq
-\frac{F}{2^*}-\frac{\beta_2}2  \td{^2F }{\xi^2} +3|A_{0,1}|^2 F =0.
\label{eqforF}
\eeq
This is the linearised equation for the modulus of $|A_{01}|$. Therefore, we may immediately spot one solution:
\beq
F_1(\xi)\propto \td{|A_{0,1}|}\xi,
\eeq
and, using that particular solution to reduce the order of (\ref{eqforF}), we find a second solution:
\beq
F_2(\xi)\propto F_1(\xi)\int_{\xi_0}^\xi\frac1{F_1(s)^2}\rd s.
\eeq
Another way to solve (\ref{eqforfm}) is to let 
\begin{align}
f_m&=\pm  G(\xi) e^{\ii m X/2^*}, &m=-\kappa\pm1,
\end{align}
then we obtain
\beq
-\frac{G}{2^*}-\frac{\beta_2}2  \td{^2G }{\xi^2} +|A_{0,1}|^2G =0.
\label{eqforG}
\eeq
One solution is, simply,
\beq
G_1\propto|A_{0,1}|.
\eeq
Again, reducing the order of (\ref{eqforG}), we find a second one:
\beq
G_2\propto G_1(\xi)\int_{\xi_0}^\xi\frac1{G_1(s)^2}\rd s.
\eeq
In the case $\beta_2<0, 2^*=2$, we obtain
\begin{align}
 F_1(\xi)&=-\frac{1}{|\beta_2|}\sech\left(\frac{\xi}{|\beta_2|^{1/2}}\right)\tanh\left(\frac{\xi}{|\beta_2|^{1/2}}\right),\\
 F_2(\xi)&=\frac{5|\beta_2|^{3/2}}{4}F_1(\xi)\bigg[6\frac{\xi-\xi_0}{|\beta_2|^{1/2}}
+ \sinh\left(\frac{2\xi}{|\beta_2|^{1/2}}\right)\non\\
&-4\coth\left(\frac{\xi}{|\beta_2|^{1/2}}\right)\bigg],\\
G_1(\xi)&=\frac{-1}{|\beta_2|^{1/2}}\sech\left(\frac{\xi}{|\beta_2|^{1/2}}\right),\\
G_2(\xi)&=\frac{-3|\beta_2|^{3/2}}{2}G_1(\xi)\bigg[\frac{\xi-\xi_0}{|\beta_2|^{1/2}}+\frac12\sinh\left(\frac{2\xi}{|\beta_2|^{1/2}}\right)\bigg].
\end{align}
On the other hand if $\beta_2>0, 2^*=1$, we obtain
\begin{align}
 F_1(\xi)&=\frac{1}{|\beta_2|}\sech^2\left(\frac{\xi}{|\beta_2|^{1/2}}\right),\\
 F_2(\xi)&=\frac{5|\beta_2|^{3/2}}{4}F_1(\xi)\bigg[\frac{3}{2}\frac{\xi-\xi_0}{|\beta_2|^{1/2}}
+ \sinh\left(\frac{2\xi}{|\beta_2|^{1/2}}\right)\non\\&
+\frac{1}{8}\sinh\left(\frac{4\xi}{|\beta_2|^{1/2}}\right)\bigg],\\
G_1(\xi)&=\frac{-1}{|\beta_2|^{1/2}}\tanh\left(\frac{\xi}{|\beta_2|^{1/2}}\right),\\
G_2(\xi)&=-3|\beta_2|\left[\frac{\xi-\xi_0}{|\beta_2|^{1/2}}\tanh\left(\frac{\xi}{|\beta_2|^{1/2}}\right)-1\right].
\end{align}
With that particular choice of integration constants, the above solutions have the following asymptotic behaviours in the vicinity of $\xi_0$:
\begin{align}
F_1&\sim\frac{i^*}{\left(\xi-\xi_0\right)^2},&
F_2&\sim i^*\left(\xi-\xi_0\right)^3,\\
G_1&\sim\frac{i^*}{ \xi-\xi_0  },&
G_2&\sim i^* \left(\xi-\xi_0\right)^2
\end{align}
Out of these asymptotic behaviours, only that of $F_2$ is compatible with (\ref{innersol1}). The other solutions connects with late terms in the vicinity of $\xi_0$ that correspond to smaller values of $\alpha$ and which are therefore subdominant in the large-$l$ limit. Matching with  (\ref{innersol1}), we thus find, away from $\xi=\xi_0$ that
\beq
E_{l}\sim \frac{\Gamma(l+4)F_2(\xi)}{\left(\xi_0-\xi\right)^{l+4}}
\sum_{\kappa}\sum_{m=-\kappa\pm1} \lambda_\kappa \left(\ii/\kappa\right)^l  e^{\ii m (x-t+ X/2^*)}
\label{outersol0}
\eeq
In this sum, the contributions associated to $\kappa=\pm2$ dominate as $l\to\infty$ and are therefore the only ones that matter to our discussion. So far, we have  only been treating the singularity $\xi=\xi_0$. Additional terms arise from the singularity at $\xi=-\xi_0$. They can be deduced from the former ones by the fact that the solution must be real where $x$, $t$, and $\xi$ are real. Eventually, we obtain
\begin{multline}
E_{l}\sim \frac{\Gamma(l+4)F_2(\xi)}{\left(\xi_0-\xi\right)^{l+4}} 
\sum_{\kappa}\sum_{m=-\kappa\pm1} \lambda_\kappa \left(\ii/\kappa\right)^l  e^{\ii m (x-t+ X/2^*)}\\
+\frac{\Gamma(l+4)\bar F_2(\xi)}{\left(-\xi_0-\xi\right)^{l+4}} 
\sum_{\kappa}\sum_{m=-\kappa\pm1} \lambda_\kappa \left(-\ii/\kappa\right)^l  e^{\ii m (x-t+ X/2^*)}
\label{outersol1}
\end{multline}
Note that the far field behaviour of $F_2$ as $\xi\to\infty$ is
\beq
F_2(\xi)\sim
\left\{
\begin{matrix}
-(5/4)|\beta_2|^{1/2} e^{\xi/|\beta_2|^{1/2}}&\text{  if \quad}\beta_2<0,\\
(5/16)|\beta_2|^{1/2} e^{2\xi/|\beta_2|^{1/2}}&\text{  if \quad}\beta_2>0,
\end{matrix}
\right.
\label{farfieldR2}
\eeq

\subsection{Location of the Stokes line}
Following Dingle~\cite{Dingle-1973}, a Stokes line occurs in the region of the complex-$\xi$ plane where successive terms have the same phase. Here, there is  a Stokes line (or ray) emanating from the singularity $\xi_0$. In the first line of (\ref{outersol1}) on passing from one order of the calculation to the next, a factor  $\ii/\kappa(\xi_0-\xi)$ is applied. This factor must be a positive real number, say $1/\mu$. Thus the Stokes ray is given by
\begin{align}
\xi&=\xi_0-\ii\mu/\kappa,
&\mu\in \mathbb{R}_0^+.
\end{align}
The Stokes ray points in the downwards direction if $\kappa>0$ and upwards if $\kappa<0$. Hence, from the singularity $\xi_0$, only the ray with $\kappa=2$ will cross the real axis and generate a new contribution to the solution there. Conversely, with  the singularity $-\xi_0$, located below the real axis, one should only count the late terms of the series associated to $\kappa=-2$. [\textit{With a quadratic nonlinearity, the same discussion holds with the substitution $\kappa=\pm2\to\pm1$.}]

\section{Truncating the system}
From what precedes, each new terms of the series expansion (\ref{expansion}) gets smaller by a factor $\eps$ but larger by a factor $(l+4)/\kappa|\xi_0-\xi|$. At a given distance $r=|\xi_0-\xi|$ from the top singularity, optimal truncation thus happens at order $L-1$, where $L=\lfloor|\kappa r|/\eps\rfloor$. We define
\beq
E^{(L-1)}=\sum_{l=0}^{L-1}\eps^lE_l,
\eeq
so that
\beq
E= E^{(L-1)} +R,
\label{L-1truncation}
\eeq
and we must now derive an equation for the remainder $R$. An estimation of the size of $R$ is given by $\eps^LE_L$, which is exponentially small in $\eps$. Hence, we may safely neglect $R^2$ compared to $R$ and anticipate that $R$ satisfies a linearized version of (\ref{wave2}) plus forcing terms:
\begin{multline}
\pd{}{x^2}R+\int_{-\infty}^\infty\beta^2(\omega)\hat R (x,\omega)e^{-\ii\omega t}\rd \omega +2\eps^2 E_0^2 R
\\
\sim \text{rhs}^{+}
+\text{rhs}^{-},
\label{remeq1}
\end{multline}
where $\text{rhs}^{\pm}$ stem from the truncation of the asymptotic series (\ref{expansion}) at order $L-1$. More specifically, there are two contributions:  $\text{rhs}^{+}$, associated to the singularity $\xi_0$ of the leading-order soliton approximation, and  $\text{rhs}^{-}$,  associated to the singularity at $-\xi_0$ in that same approximation. All of the intricate foregoing calculations precisely aimed at deriving $\text{rhs}^{+}$ and $\text{rhs}^{-}$, which is what we are about to do. But first, let us note that, by the linearity of Eq.~(\ref{remeq1}), we may write  $R$ as the sum of  particular solutions 
\beq
R=R^{+}+R^{-}
\eeq
associated  to $\text{rhs}^{+}$ and $\text{rhs}^{-}$, respectively. Once $R^{+}$ has been calculated, $R^{-}$ can be deduced by the fact that $R^{+}+R^{-}$  is real on the real axis. We may therefore restrict our attention to $\text{rhs}^{+}$, and hence, on (\ref{outersol0}) to derive it.

The various terms appearing in $E^{(L-1)} $ are such that all the $O(\eps^l)$ terms up to $l=L-1$ vanish when substituting (\ref{L-1truncation})  in Eq.~(\ref{wave2}). Let us therefore focus on the terms of order $\eps^L$ and higher. We have
\begin{multline}
\pd{}{x^2}E^{(L-1)}\sim\ldots -2\eps\pd{^2}{x\dd\xi}\eps^{L-1}E_{L-1} \\ +\eps^2\pd{^2}{\xi^2}\left(\eps^{L-1}E_{L-1} +\eps^{L-2}E_{L-2}\right)\\
=\ldots
+\sum_me^{\ii m(x-t)}\left[-2\ii m\eps\pd{}{\xi}\eps^{L-1}A_{L-1,m}\right.\\
\left.
+\eps^2\pd{^2}{\xi^2}\left(\eps^{L-1}A_{L-1,m}+\eps^{L-2}A_{L-2,m}\right)
\right]
\label{d2E(L-1)/dx2}
\end{multline}
where we omitted to write all terms up to order $L$ and kept only the dominant remaining ones. Similarly,
\begin{multline}
\int_{-\infty}^\infty\beta^2(\omega)\hat E^{(L-1)} (x,\omega)e^{-\ii\omega t}\rd \omega
\sim\ldots\\
+\sum_m e^{\ii m(x-t)} \sum_{q\geq1}\sum_{n\geq q}\Lambda^m_n\left(\ii\eps\dd/\dd\xi\right)^n\eps^{L-q}A_{L-q,m}
\label{truncateddispersion}
\end{multline}
Next, the nonlinear terms are
\begin{multline}
\frac{2\eps^2}3\left(E^{(L-1)}\right)^3
\sim\ldots + 2\eps^2E_0^2\sum_{q=1}^2\eps^{L-q}E_{L-q}.
\end{multline}
Now, near optimal truncation, all terms $\eps^{L-q}E_{L-q}$ with $q=O(1)$ approximately have the same magnitude. Moreover, the operator $\eps\dd/\dd\xi$, when applied to a function of the form (\ref{outeransatz}) yields a contribution that is proportional to $\eps(l+4)$ which is $O(1)$ when $l=O(L)$. As a result, the nonlinear terms just derived are a factor $\eps^2$ smaller than those in (\ref{d2E(L-1)/dx2}) and in (\ref{truncateddispersion}) and can be neglected in comparison. We thus have
\begin{multline}
\text{rhs}^{+}+\text{rhs}^{-}=\sum_m e^{\ii m(x-t)}\Bigg[
2\ii m\eps\pd{}{\xi}\eps^{L-1}A_{L-1,m}   \\
-\eps^2\pd{^2}{\xi^2}\left(\eps^{L-1}A_{L-1,m} +\eps^{L-2}A_{L-2,m}\right)\\
\left.
-\sum_{q\geq1}\sum_{n\geq q}\Lambda^m_n\left(\ii\eps\dd/\dd\xi\right)^n\eps^{L-q}A_{L-q,m}
\right]
\label{rhs:L-1}
\end{multline}
Focusing on $\text{rhs}^{+}$, that is on (\ref{outersol0}), the terms  with $\kappa=2$, \textit{i.e.} the harmonics  $m=-1,-3$, dominate upon crossing the Stokes line, so we ignore the others. [\textit{With a quadratic nonlinearity, $\kappa=1$, $m=-1,0$.}] With $\kappa=2$, each successive term is identical in phase and amplitude, up to an $O(1/L)$ difference on the Stokes line: $\eps^{L-q}A_{L-q,m}\sim\eps^{L-1}A_{L-1,m}$ for all integers $q$ of order one. $\text{rhs}^{+}$ thus simplifies as
\begin{multline}
\text{rhs}^{+}
\sim
\sum_{m=-\kappa\pm1}e^{\ii m(x-t)}\left[
2\ii\eps\pd{}{\xi}\left(
m +\ii \eps\pd{}{\xi}
\right) \right.\\
\left.
-\sum_{q\geq1}\sum_{n\geq q}\Lambda^m_n\left(\ii\eps\dd/\dd\xi\right)^n\right]
\eps^{L-1}A_{L-1,m},
\label{remeq2}
\end{multline}
with $\kappa=2$. Regarding the double sum above, we note the following identity:
\beq
\sum_{q\geq1}\sum_{n\geq q}a_n = \sum_{n\geq1}n a_n,
\eeq
provided the right hand side exists. Therefore,
\begin{multline}
\text{rhs}^{+}
\sim
\sum_{m=-1,-3}e^{\ii m(x-t)}\left[
2\ii\eps\pd{}{\xi}\left(
m  
+\ii\eps\pd{}{\xi}
\right) \right.\\
\left.
-\sum_{n\geq 1} n \Lambda^m_n\left(\ii\eps\dd/\dd\xi\right)^n\right]
\eps^{L-1}A_{L-1,m}.
\label{remeq3}
\end{multline}
Further, and again with $O(1/L)$ accuracy, $\dd/\dd\xi\sim L/(\xi_0-\xi)$:
\begin{multline}
\text{rhs}^{+}
\sim
\frac{\ii\eps L}{\xi_0-\xi}\sum_{m=-1,-3}e^{\ii m(x-t)}\left[2
\left(
 m  
+\frac{\ii \eps L}{\xi_0-\xi}
\right) \right.\\
\left.
-\sum_{n\geq 1} n \Lambda^m_n\left(\frac{\ii\eps L}{\xi_0-\xi}\right)^{n-1}\right]
\eps^{L-1}A_{L-1,m}.
\label{remeq4}
\end{multline}

\subsection{Local behaviour near the Stokes line}
Berry showed on some examples that exponentially small terms are switched on not discontinuously but in an $O(\eps^{1/2})$-thin region comprising the Stokes line~\cite{Berry-1988}. This observation has been confirmed in many instances~\cite{Berry-1988,Chapman-1998,Chapman-2006,Kozyreff-2006,Chapman-2009,Joshi-2017}. Based on this knowledge, let us write 
\beq
\xi=\xi_0-\ii r+\eps^{1/2}s.
\label{xi2s}
\eeq
In the following, it will sometimes be convenient to write the asymptotically equivalent expression
\begin{equation}
\xi_0-\xi=\ii r-\eps^{1/2}s 
\sim\ii re^{\ii\eps^{1/2}s/r+\frac12\eps(s/r)^2}
\label{Stokesvariable}
\end{equation}
In (\ref{remeq4}), we have, using (\ref{xi2s}),
\begin{multline}
2\left( m +\frac{\ii \eps L}{\xi_0-\xi}\right) 
-\sum_{n\geq 1} n \Lambda^m_n\left(\frac{\ii\eps L}{\xi_0-\xi}\right)^{n-1}\\
\sim
2 \left( m  +\kappa-\ii\eps^{1/2}\frac{\kappa s}{r} \right) 
-\sum_{n\geq 1} n \Lambda^m_n\left(\kappa-\ii\eps^{1/2}\frac{\kappa s}{r}\right)^{n-1}\\
\sim
2 \left( m  +\kappa-\ii\eps^{1/2}\frac{\kappa s}{r} \right) \\
-\sum_{n\geq 1} \Lambda^m_n \left( n\kappa^{n-1}-n(n-1)\kappa^{n-2}\ii\eps^{1/2}\frac{\kappa s}{r}\right)\\
=2 \left( m  +\kappa-\ii\eps^{1/2}\frac{\kappa s}{r} \right) -\td{}{\kappa}\beta^2(m+\kappa)\\
+\ii\eps^{1/2}\frac{\kappa s}{r}\td{^2}{\kappa^2}\beta^2(m+\kappa)\\
=2\ii\beta_2 \eps^{1/2}\frac{\kappa s}{r}
\end{multline}
Next, we must evaluate $\eps^{L-1}A_{L-1,m}$ in the vicinity of the stokes line. With (\ref{Stokesvariable}) and using $\Gamma(z)\sim\sqrt{2\pi/z}\left(z/e\right)^z$, we have, in (\ref{outersol0}), with $l=L-1\sim\kappa r/\eps$,
\begin{multline}
\frac{\eps^{L-1}\Gamma(L+3) }{\left(\xi_0-\xi\right)^{L+3}}\left(\ii/\kappa\right)^{L-1}\\
\sim
\left(\kappa/\eps\right)^4\sqrt{\frac{2\pi}{L+3}}\left(\frac{\ii\eps(L+3)}{e \kappa (\xi_0-\xi)}\right)^{L+3}\\
\sim
\left(\kappa/\eps\right)^4\sqrt{\frac{2\eps\pi}{\kappa r}}   e^{-\kappa /\eps(r+ \ii\eps^{1/2} s)- \frac12\kappa   s^2/r}\\
= \left(\kappa/\eps\right)^4\sqrt{\frac{2\eps\pi}{\kappa r}}   e^{\ii \kappa /\eps(\xi_0-\xi)- \frac12\kappa   s^2/r}\\
= \left(\kappa/\eps\right)^4\sqrt{\frac{2\eps\pi}{\kappa r}}  e^{\ii\kappa(x-t-x_0+X/2^*)} e^{- |\kappa \Im(\xi_0)| /\eps}e^{- \frac12\kappa   s^2/r}.
\end{multline}
Hence, we obtain
\begin{multline}
\pd{}{x^2}R^{+}+\int_{-\infty}^\infty\beta^2(\omega)\hat R^{+} (x,\omega)e^{-\ii\omega t}\rd \omega+2\eps^2 E_0^2 R^{+}
\\
\sim
2\ii\eps \beta_2 \lambda_\kappa F_2(\xi)\left(\kappa/\eps\right)^4
e^{-|\kappa \Im(\xi_0)| /\eps}\sqrt{\frac{2\kappa\pi}{ r}} \, e^{-i\kappa x_0}\\
\sum_me^{\ii (m+\kappa)(x-t+X/2^*)} 
 \frac{\kappa s}{r}e^{- \frac12\kappa   s^2/r}.
\label{remeq5}
\end{multline}
We thus see that the right hand side of (\ref{remeq5}) varies very rapidly, on the $s$-scale, becoming negligible as soon as $s^2$ exceeds a few times $r/\kappa$. On the $\xi$-scale, this is an $O(\eps^{1/2})$-thin region. In the vicinity of the Stokes line, we let
\begin{multline}
R^{+}\sim 
4\ii\pi  \lambda_\kappa F_2(\xi)\left(\kappa/\eps\right)^4
e^{-| \kappa \Im(\xi_0)| /\eps}e^{-\ii\kappa x_0}\\
\times
\sum_me^{\ii (m+\kappa)(x-t+X/2^*)} S_m(r,s)
\end{multline}
with $r$ fixed and $s$ related to $x$ and $t$ through (\ref{Stokesvariable}) and (\ref{slowscales}). Substitution in (\ref{remeq5}) yields
\begin{multline}
-\left(m+\kappa\right)^2 S_m-2\ii\left(m+\kappa\right)\eps^{1/2}\pd{S_m}{s}+\eps\pd{^2S_m}{s^2}\\
+\beta^2(m+\kappa+\ii\eps^{1/2}\dd/\dd s)S_m\\
\sim -\eps\beta_2\pd{^2S_m}{s^2}\sim\eps \beta_2 \sqrt{\frac{\kappa}{ 2\pi r}} 
 \frac{\kappa s}{r}e^{- \frac12\kappa   s^2/r}.
\end{multline}
The solution of this last equation is
\beq
S_m(r,s)=\frac1{\sqrt\pi}\int_{-\infty}^{\sqrt{\kappa/(2r)}s}e^{-u^2}\rd u
\eeq
With this solution, $R^{+}\to0$ as $s\to-\infty$, and 
\begin{multline}
R^{+}
\to 8\ii\pi  \lambda_\kappa F_2(\xi)\left(\frac\kappa \eps\right)^4
e^{-| \kappa \Im(\xi_0)| /\eps}\\
\times e^{-\ii\kappa x_0} 
\cos\left(x-t+X/{2^*}\right) 
\end{multline}
as $s\to\infty$. Hence, upon crossing the Stokes line that joins $\xi_0$ and $-\xi_0$, one obtains the contribution
\begin{multline}
R^{+}+R^{-}\sim16\pi\kappa^4\lambda_\kappa\eps^{-4} e^{-\kappa\Im(\xi_0) /\eps}\\
\times\Re\left( \ii F_2(\xi)
e^{-i \kappa x_0} \right)
\cos(x-t+X/2^*) 
\label{theremainder}
\end{multline}
to the remainder $R$.

\section{Soliton equation of motion}
Let us first consider the bright soliton case. Adding the leading order expression of the slowly accelerating soliton, see Eq.~(\ref{accelerate1}), and the remainder, Eq.~(\ref{theremainder}), we finally obtain, as $\xi\to\infty$ and with $c$ exponentially small
\begin{multline}
E\sim \cos\left(x-t+X/2\right) e^{\xi/|\beta_2|^{1/2}}\\
\times \left(-\td{c}{X}-20\pi\kappa^4|\beta_2|^{1/2}\lambda_\kappa\eps^{-4}e^{-\kappa\Im(\xi_0) /\eps}\sin\kappa x_0\right),
\label{motionbright}
\end{multline}
where we have used the large-$\xi$ asymptotic expressions (\ref{farfieldRabright}) for $R_a$ and (\ref{farfieldR2}) for $F_2$, and where $\zeta\sim\xi/|\beta_2|^{1/2}$ on account of the exponential smallness of $c$.
In order to prevent unacceptable exponential divergence, we must set the factor between parentheses to zero. This is the sought-after result, since $\dot x_0=\eps v_p c$ and $\dot c=\eps^2\beta_0v_p\td cX$. Eventually, in the original variables, and reverting to  $\veps$ as the small parameter, we obtain
\beq
\ddot x_0+\eta\frac{v_l^2}{4}\kappa\beta_0 \sin\left(\kappa \beta_0 x_0\right)=0
\eeq
with
\begin{align}
 v_l^2&=80\pi\kappa^3|\lambda_\kappa| \veps^{-1} e^{-\kappa\pi /2\veps}v_p^2,
&\eta&=\sign(\lambda_\kappa).
\end{align}

Secondly, we consider dark solitons ($\beta_2>0$). Here, as $\xi\to\infty$,
\begin{multline}
E\sim 0.25 \cos\left(x-t+X\right)  e^{2\xi/|\beta_2|^{1/2}}  \\
\times \left(\td{c}{X}+20\pi\kappa^4|\beta_2|^{1/2}\lambda_\kappa \eps^{-4}e^{-\kappa\Im(\xi_0) /\eps}\sin\kappa x_0\right),
\label{motiondark}
\end{multline}
where we used the far-field expressions (\ref{farfieldRadark}) and (\ref{farfieldR2}) for $R_a$ and $F_2$, respectively. We thus obtain exactly the same equation of motion as with the bright soliton.

\section{Conclusions and perspectives}
The law of propagation of  wave packets  at the group velocity is one of the most fundamental in physics, given its simplicity and scope of application. This paper brings an exception to the rule, in the case where phase and group velocities are very close. In an exponentially small range of parameter, the nonlinear Schr\"odinger equation becomes invalid as far as soliton dynamics is concerned. There, the motion of bright and dark solitons is locally equivalent to that of a pendulum. To obtain this rather simple-looking result, it was necessary to carry out a calculation beyond all orders of the classical multiple scales expansion that underlies the NLSE. We have treated a general class of wave equations that describes many physical situations, which gives confidence in the generality of our findings. What is required is a weak nonlinearity and the existence of an extremum of the phase velocity. The weak nonlinearity, which exists in almost any classical physical system, leads to the existence of harmonics of the fundamental wave. These  harmonics interact very weakly with the fundamental one to produce the pinning force. The higher the separation between harmonics, the weaker the interaction. This explains why quadratic nonlinearities lead to stronger pinning forces than cubic ones, for the later only lead to odd harmonics of the fundamental signal.

The ultimate result of the present beyond-all-orders calculation, Eqs.~(\ref{motionbright}) and (\ref{motiondark}), correspond to the smallest possible absolute value of $\kappa$, \textit{i.e.} to the closest nonlinear harmonics of the fundamental ones. The other harmonics, associated to larger absolute values of $\kappa$, contributes in principle to the equation of motion too, even though they are exponentially smaller than first term. This suggests that the set of all these contributions would form a trans-series.

It would be desirable to numerically demonstrate the dynamics described here but this is a challenging task. One should indeed simulate the complete wave model Eq.~(\ref{wave2}), including both the fast oscillations of the carrier wave and their slow envelope at the same time, since the dynamics is precisely governed by the interplay between these widely separated spatio-temporal scales. Moreover, one should let the field evolve over distances or durations that are exponentially larger than the elementary oscillations, since the characteristic speed of the soliton centre of mass relative to the carrier wave is given by the exponentially small $v_l$ in the region of interest. This requires an efficient and stable numerical algorithm. An attempt was made in the course of this research but it was unsuccessful: our numerical code with explicit time-stepping was prone to  numerical instabilities as soon as the nonlinearity was present. As a result, soliton pulse evolution could be monitored for only about 40 time units and $\varepsilon$ was constrained to be no larger than 0.1. This was not enough to confirm the phase portrait  of Fig.~\ref{fig:pendulum}, because the acceleration and deceleration of the soliton takes place over an $O\left(1/v_l\right)$ characteristic time. This is a similar situation to the one encountered with localized patterns pushed by distant boundaries~\cite{Kozyreff-2009b,Kozyreff-2011}, where the slow dynamics requires numerical integration over tens of thousand or even millions of unit time. In the absence of  a proper numerical investigation, let us stress that a partial numerical confirmation of  Eq.~(\ref{eq:central}) already exists: its  stationary solutions and their stability are consistent with the numerical results obtained in the particular case of~\cite{YangAkylas-1997,Calvo-2000}.

Beside challenging the conventional view of wave propagation at the fundamental level, the results of this paper may find their way to application. One example is soliton-based optical frequency combs~\cite{Hansson-2013,Chembo-2013}. In this frame, the phase of wave packets (light pulses) must be tightly controlled, and it is precisely that phase which is governed by Eq.~(\ref{eq:central}). In the same vein, nonlinear optical pulses can travel over very long distances compared to their width in optical fibers with low attenuation. Over such distances,  exponentially small effects, if present, may have time to qualitatively affect the propagation of wave packets.

\acknowledgments
G.K. is a Research Associate of the Fonds de la Recherche Scientifique - FNRS (Belgium.) This research started on the occasion of the Workshop ``Nonlinear in optics: theory and experiments'', Besan\c{c}on, 4-5 November 2015. 

\appendix

\section{Derivation of the wave equation~(\ref{wave2})}
\label{appendix:eq1}

\subsection{Electromagnetic waves} 
In a non magnetic medium, the electric and magnetic fields, respectively $\ve \EE$ and $\ve \HH$ are governed by~\cite{Jackson-1999},
\begin{align}
\curl\boldsymbol\HH&=\pd{\boldsymbol\DD}t+\pd{\boldsymbol\PP_{nl}}t,
& \curl\boldsymbol\EE&=-\mu_0\pd{\boldsymbol\HH}t,
\end{align}
where the displacement field is separated into a contribution $\boldsymbol\DD$ that is a linear functional of $\boldsymbol\EE$, and a nonlinear polarisation term $\boldsymbol\PP_{nl}$. By cross differentiation and using the fact that the fields are divergence free, we may rewrite the above two equations as
\beq
\boldsymbol{\nabla}^2\boldsymbol\EE-\mu_0\pd{^2\boldsymbol\DD}{t^2}=\mu_0\pd{^2\boldsymbol\PP_{nl}}{t^2}.
\eeq
Let us now particularise the equation to an electric field with a single component in the $y$ direction, and an input amplitude $E_0$:  $\boldsymbol\EE= E_0  E(x,t)\ve{\hat y}$. Then, with 
\beq
-\mu_0\pd{^2\boldsymbol\DD}{t^2}=\int_{-\infty}^\infty\beta^2(\omega) \boldsymbol{\hat\EE}(x,\omega)e^{-\ii\omega t}\rd \omega
\label{displacement}
\eeq
 and $\boldsymbol\PP_{nl}=\varepsilon_0\chi^{(3)}|\boldsymbol\EE|^2\boldsymbol\EE$ (see for instance \cite{Agrawal-book,Newell-book}), we get 
\beq
\pd{^2E}{x^2}+\int_{-\infty}^\infty\beta^2(\omega)\hat E(x,\omega)e^{-\ii\omega t}\rd \omega
=\frac{\chi^{(3)}|E_0|^2}{c^2} \pd{^2}{t^2}\left(E^3\right).
\eeq
Hence,  we recover (\ref{wave2}) and the nonlinearity (\ref{wavenonlinearity}) with
\beq
\eps^2 = \frac{3\chi^{(3)}|E_0|^2\omega_0^2}{2\beta(\omega_0)^2c^2}\ll1.
\eeq
If, instead of free-space, one considers wave propagation in a waveguide, then we may approximately write $\boldsymbol{\mathcal E}=  E(x,t) \boldsymbol{\Phi}(\ve r_\perp)$, where $\boldsymbol{\Phi}(\ve r_\perp)$ is the vector distribution of a waveguide mode, which depends on the transverse coordinates $\ve r_\perp$  (see, e.g.~\cite{Snyder-1984}). Then Eq.~(\ref{displacement}) can again be used, provided that $\beta(\omega)$ is the dispersion function of the waveguide and not that in free space. In general  $\boldsymbol{\Phi}(\ve r_\perp)$ also depends on frequency but for a wavepacket, this dependence can safely be neglected.

\subsection{Elastic waves}
Let us next consider waves propagating along an elastic beam. There, the vertical displacement $w(x,t)$ satisfies~\cite{Howell-2009}
\beq
\rho\pd{^2w}{t^2}-T \pd{^2w}{x^2}+B\pd{^4w}{x^4}=f,
\eeq
where $T$ is the tension in the beam, $B$ its bending stiffness and $f$ is a distributed force, which may depend nonlinearly on $w$. Let us assume a nonlinear restoring force $f=- f_1w-f_2w^2 $. Then by way of using the spatial Fourier transform of $w$,
\beq
\hat w(k,t)=\frac1{2\pi}\int_{-\infty}^\infty w(x,t) e^{-ikx}\rd x,
\eeq
the beam equation can be rewritten as
\beq
\pd{^2w}{t^2}+\int_{-\infty}^{\infty}\omega^2(k)\hat w(k,t) e^{ikx}\rd k+f_2w^2/\rho=0,
\label{eq:beam}
\eeq
with $\omega^2(k)=(Tk^2+Bk^4+f_1)/\rho$. This is of the same form as (\ref{wave2}), except for the exchange of coordinate $x$ and $t$ and the replacement $\beta(\omega)\to\omega(k)$. A minimum of $v_p$ is obtained at the wave number  $k_0=(f_1/B)^{1/4}$. Note in particular that if the constants $B$, $T$, $f_1$, $f_2$, and $\rho$ are chosen such that the equation reads $\dd{^2w}/\dd{t^2}+\frac12\left(w+\dd^4w/\dd x^4\right)+\sqrt{9/38} \,\eps w^2=0$, then the critical wave and angular frequency are both unity and the NLSE reads $i\dd A/\dd T+\dd^2A/\dd\xi^2+|A|^2A=0$, where $T=\eps^2t$.

\subsection{Plasma waves}
Finally, let us consider the model for cold plasmas used  by Taniuti et al~\cite{Taniuti-1968}
\begin{align}
\pd{n}t+\pd{\left(nu\right)}{x}&=0,\\
\left(\pd{}t+u\pd{}{x}\right)u+n^{-1}\pd{}x\left(\frac12|\mathcal B|^2\right)&=0,
\end{align}
\beq
\left(\pd{}t+u\pd{}{x}\right)\left(\mathcal V+iR_e^{-1}n^{-1}\pd{\mathcal B}x\right)-n^{-1}\pd{\mathcal B}x =0,
\label{cold1}
\eeq
\begin{multline}
\pd{}t\left(\mathcal B-iR_i^{-1}\pd{\mathcal V}x\right)-\pd{\mathcal V}{x}=\\
-\pd{}x\left[u\left(\mathcal B-iR_i^{-1} \pd{\mathcal V}x\right)\right] ,
\label{cold2}
\end{multline}
\begin{align}
\mathcal V&=v-iw,
& \mathcal B&=B_y-iB_z,
\end{align}
where $n$ is the density, $(u,v,w)$ is the cartesian velocity field, $B_y$ and $B_z$ are the transverse components of the magnetic field, with a constant applied magnetic field in the $x$ direction. Finally, $R_e$ and $R_i$ are normalised cyclotron frequencies for the electrons and ions, respectively. Assuming $n=1-\eps\hat n$ and $u=\eps \hat u$, Eqs.~(\ref{cold1}) and (\ref{cold2}) read
\begin{align}
\pd{}t\left(\mathcal V+iR_e^{-1}\pd{\mathcal B}x\right)-\pd{\mathcal B}x &=\eps f_1(\hat u,\hat n,\mathcal B,\mathcal V),\label{cold3}\\
\pd{}t\left(\mathcal B-iR_i^{-1} \pd{\mathcal V}x\right)-\pd{\mathcal V}{x}&=\eps f_2(\hat u,\hat n,\mathcal B,\mathcal V)\label{cold4}
\end{align}
Defining $\mathcal F=\mathcal V+iR_e^{-1}\pd{\mathcal B}x$ and $\mathcal G=\mathcal B-iR_i^{-1} \pd{\mathcal V}x$, we have, using (\ref{cold3}) and (\ref{cold4}), $\mathcal V\sim\mathcal F-iR_e\pd{\mathcal F}{t} $ and $\mathcal B\sim\mathcal G+iR_i^{-1}\pd{\mathcal G}{t}$ in the small-$\eps$ limit. Furthermore, let us introduce a new function $\mathcal H$ as 
\beq
\mathcal H = \left(1+iR_i^{-1}\pd{}{t}\right)\left(1-iR_e^{-1}\pd{}{t}\right)\mathcal F. 
\eeq
Then Eqs.~(\ref{cold3}) and (\ref{cold4}) can be combined to yield 
\beq
\pd{^2\mathcal H}{x^2}+\int_{-\infty}^\infty\frac{\omega^2\mathcal{\hat H}(x,\omega)e^{-i\omega t}}{\left(1-iR_i^{-1}\omega\right)\left(1+iR_e^{-1}\omega\right)}\rd \omega=\eps N,
\eeq
where $N$ is a nonlinear functional of $\mathcal H$.

\section{Derivation of the functions $g_q(l)$}\label{appendix:gq(l)}
In this section, we derive the expressions of the functions $g_q(l)$ appearing in (\ref{factor}). Omitting the label $m$, we have
\beq
\sum_{q=0}^n \frac{n!  g_q(l)}{(n-q)!}=
\frac{l!\Gamma(l+\alpha-n)(\ii/\kappa)^{n-l}B_{l-n}}{(l-n)!\Gamma(l+\alpha)}
\label{app:q:1}
\eeq
where
\beq
(\ii/\kappa)^{n-l}B_{l-n}\sim b^{(0)}+(l-n)^{-1}b^{(1)}+(l-n)^{-2}b^{(2)}+\dotsb
\eeq
Evaluating (\ref{app:q:1}) with $n=0$, we directly obtain
\beq
g_0(l)=(i/\kappa)^{-l}B_{l}\sim b^{(0)}+l^{-1}b^{(1)}+l^{-2}b^{(2)}+\dotsb
\eeq
For $n\geq1$, the $g_n(l)$ are obtained by recurrence:
\beq
g_n(l)=
\frac{l!\Gamma(l+\alpha-n)(\ii/\kappa)^{n-l}B_{l-n}}{n!(l-n)!\Gamma(l+\alpha)}
-\sum_{q=0}^{n-1} \frac{g_q(l) }{(n-q)!},
\eeq
which, upon expansion in the large-$l$ limit yields (\ref{g0})-(\ref{g2}).

%


\end{document}